# Adopting Softer Approaches in the Study of Repository Data: A Comparative Analysis

Sherlock A. Licorish and Stephen G. MacDonell

*SERL, School of Computing and Mathematical Sciences*
*AUT University, Auckland 1142, New Zealand*
*slicoris@aut.ac.nz, smacdone@aut.ac.nz*

## ABSTRACT

*Context:* Given the acknowledged need to understand the people processes enacted during software development, software repositories and mailing lists have become a focus for many studies. However, researchers have tended to use mostly mathematical and frequency-based techniques to examine the software artifacts contained within them. *Objective:* There is growing recognition that these approaches uncover only a partial picture of what happens during software projects, and deeper contextual approaches may provide further understanding of the intricate nature of software teams' dynamics. We demonstrate the relevance and utility of such approaches in this study. *Method:* We use psycholinguistics and directed content analysis (CA) to study the way project tasks drive teams' attitudes and knowledge sharing. We compare the outcomes of these two approaches and offer methodological advice for researchers using similar forms of repository data. *Results:* Our analysis reveals significant differences in the way teams work given their portfolio of tasks and the distribution of roles. *Conclusion:* We overcome the limitations associated with employing purely quantitative approaches, while avoiding the time-intensive and potentially invasive nature of field work required in full case studies.

## 1. INTRODUCTION

Motivated by the need to improve software project performance and project success rates [1], there has been growing interest in examining software repository data to provide understandings for the ways software teams co-exist during the provision of software solutions [2]. Exploring software process issues through such an approach has become popular given the ease with which artifacts are extracted and the unobtrusive nature of investigating issues from these sources. Most of the works examining communication artifacts, such as electronic messages, change request histories and blogs, to explain teams' behavioral processes have tended to employ frequency-based analysis techniques (e.g., Social Network Analysis (SNA)). While these techniques enable the detection of certain patterns and provide partial accounts for the ways in which software teams work, reservations have been expressed regarding the effectiveness of these approaches in delivering understandings of the deeper psychological nature of team dynamics [3].

It is therefore necessary to supplement these quantitative analyses with more contextual and qualitative examinations of team interactions if we are to provide deeper understandings of the ways in which software teams operate [3]. In this study we report our experiences utilizing psycholinguistic and directed content analysis approaches in the study of repository data from two Jazz teams (based on the IBM$^R$ Rational$^R$ Team Concert$^{TM}$ (RTC)[1]). We explore the way project tasks drive team members' attitudes and knowledge sharing. Our primary intent is to demonstrate how the use of deeper analysis approaches from the social and organizational psychology disciplines illuminates unique and valuable forms of evidence around teams' behavioral processes. We provide methodological advice for software engineering (SE) research considering repository data; our outcomes could also inform software project governance.

In the next section we briefly present our theoretical background and motivation, outlining our specific research question (Section 2). We then provide our study design in Section 3. In Section 4 we present our results and analysis. Section 5 then provides our comparative summary and its implications for SE research and practice. Finally, we consider our study's limitations in Section 6.

---

[1] IBM, the IBM logo, ibm.com, and Rational are trademarks or registered trademarks of International Business Machines Corporation in the United States, other countries, or both.

## 2. BACKGROUND AND MOTIVATION

Much prior research has been dedicated to understanding the practices and processes enacted during software development with the goal of providing recommendations to improve project outcomes. In the last decade, the growing availability of software repositories has provided researchers with new avenues through which to study a range of practice-related issues, including software teams' interactions [4]. For instance, based on the study of repository data it has been shown that high levels of interaction increase individual participants' knowledge bases [5]. Other studies conducted using such sources have also found evidence of correlation between interaction frequency and defect-introducing changes [2]. Another study of team coordination [6] found mailing list activities to be dominated by just a few individuals.

In reviewing these and other related works it is evident that, while such repository-based analysis of developers' communication and interaction data provides a useful opportunity to understand the nature of software development, this subject has been approached primarily from a quantitative standpoint using mostly open source software (OSS) repositories. Issues related to poor data quality have been previously shown to plague OSS repositories [7]. There is also widespread recognition that supplementing quantitative analysis with more exploratory approaches offers avenues for outcome triangulation as well as the provision of additional insights into the software development process [8].

Accordingly, our enquiry is driven by roles theories and psychology, which show that various behaviors and traits are prevalent and necessary in some team environments or contexts, while other settings demand different attitudes for teams to succeed [9-10]. These theories may also be applicable for software engineering teams [11]. The absence of specific team attitudes and knowledge sharing behaviors may throw out team balance and challenge project success. Thus, enquiries into the ways different software teams work will deliver concrete recommendations for team composition, given teams' specific development portfolios. With the potential to learn more using deeper analysis approaches in the study of a representative software repository, and with theoretical support from roles theories and psychology [9-10], we explore:

> How do project tasks drive team members' attitudes and knowledge sharing behaviors?

## 3. STUDY DESIGN

As noted in the previous section, studies utilizing repository data in the examination of process issues have regularly employed connection frequency-based analysis approaches, providing useful but inherently partial views of the phenomena of interest. Such approaches are in line with a SE research disposition oriented towards the more technical aspects of software development activities [8]. It is now generally recognized, however, that studying these aspects in isolation places substantial limitations on the evidence that such projects can provide [8]. Additionally, there is limited feasibility of undertaking deeper contextual enquiries without engaging other research approaches as have been used and tested in other disciplines [12].

Given access to the IBM Rational Jazz repository, an environment integrating all activities related to software development – including process management, coding, issues, discussions and design – we employed an embedded case study design [13] in the analysis of two sets of project artifacts. The embedded case study approach is appropriate for understanding complex human processes by relating them to their context [13], as is the focus of the work under consideration here. We stress, however, that the primary focus of this particular study is to compare the outcomes of deeper contextual analysis techniques and to identify opportunities for triangulating results uncovered from these approaches. Thus, although our empirical results may inform software project governance and initiate requirements for software tool design, we do not claim (or intend) to present substantial and generalizable findings. During our study, we extracted Jazz and selected two cases (see subsection 3.1), and the extracted data were then analyzed using statistical techniques, linguistic analysis tools (see subsection 3.2) and directed CA (see subsection 3.3).

### 3.1 Case Repository and Data Collection

The repository examined in this work is a specific release (1.0.1) of Jazz, comprising a large amount of process data collected from development and management activities across the USA, Canada and Europe. Jazz, created by IBM, is a fully functional environment for developing software and managing the entire software development process [14]. The software includes features for work planning and traceability, software builds, code analysis, bug tracking and version control in one system. Changes to source code in the Jazz environment are only allowed as a consequence of work items (WI) being created beforehand, comprising bug reports, new feature requests or requests to enhance existing features. Jazz teams use the Eclipse-way methodology for guiding the software development process [14]. This methodology outlines iteration cycles that are six weeks in duration, comprising planning, development and stabilizing phases. Builds are executed after project iterations; also called project milestones. All information on the software process is stored in a server repository, which is accessible through a web-based or Eclipse-based RTC client interface.

We wrote a Java program to leverage the Jazz Client API to extract team information and development and

communication artifacts from the repository. These included (in addition to WIs):

- Project Workspace – each Jazz team is assigned a workspace. The workspace contains all the artifacts belonging to the specific team; also called project or team area.
- Contributors and Teams – a contributor is a practitioner who contributes to one or more features; multiple contributors form teams (led by a project manager or team leader).
- Comments or Messages – communication (enforced through Jazz itself) around WIs. Messages ranged from as short as one word (e.g., 'thanks'), to up to 1055 words (multiple pages of communication).

To support our investigation we extracted the Jazz repository and randomly selected two sets of teams' artifacts (out of 94) dedicated to solving user experience and project management tasks, respectively (see further details in Licorish and MacDonell [11]). We wrote scripts to remove all HTML tags and foreign characters from the textual data (including comments and work items descriptions).Table 1 shows that the selected project artifacts comprised 264 distinct software development tasks (WIs), involving the efforts of 123 contributors, and resulting in 1072 messages being exchanged during development.

**Table 1. Summary Statistics for Selected Projects**

| Project | Development Activities | Period (days) | Task Count | Total Messages | Total Contributors |
|---|---|---|---|---|---|
| UE | User Experience – tasks related to UI development | 304 | 54 | 460 | 33 |
| PM | Project Management – tasks under the project managers' control | 660 | 210 | 612 | 90 |
| ∑ | | 964 | 264 | 1072 | 123 |

## 3.2 Linguistic Analysis Techniques

Previous research has identified that individual linguistic style is quite stable over time and that text analysis programs are able to accurately link language characteristics to behavioral traits (see [15], for example). We employed the Linguistic Inquiry and Word Count (LIWC) software tool in our analysis of team attitudes. The LIWC is a software tool created after four decades of research using data collected across the USA, Canada and New Zealand [16]. Similar to an electronic dictionary, this tool accepts written text as input which is then processed based on the LIWC dictionary after which summarized output is provided. The different linguistic dimensions in the summary (see examples in Table 2) are said to capture the psychology of individuals by assessing the words they use [15-16]. For example, consider the following sample comment:

"We are aiming to have all the patches ready by the end of this release; this will provide us some space for the next one. Also, we are extremely confident that similar bug-issues will not appear in the future."

In this comment the author is expressing optimism that the team will succeed, and in the process finish ahead of time and with acceptable quality standards. In this quotation, the words "we" and "us" are indicators of team or collective focus, "all", "extremely", "confident" are associated with certainty, while the words "some" and "appear" are indicators of tentative processes. Words such as "bug-issues" and "patches" are not included in the LIWC dictionary and would not affect the context of its use - whether it was to indicate a fault in software code or a problem with one's immunity to, and treatment for, a disease. Although these omissions may be interpreted as representing a confounding factor due to contextual ignorance, we in fact know that the context is software development; and what is of interest, and is being captured by the tool, is evidence of attitudes. Previous work has also provided confirmation for the utility of the LIWC dictionary for examining software developers' attitudes [11]. In the current work we examine teams' attitudes via their messages, along multiple linguistic categories; Table 2 details the categories selected with theoretical justification for their inclusion.

## 3.3 Directed Content Analysis (CA)

We also studied the messages contributed by the members of the two projects using a directed CA approach, employing a hybrid classification scheme adapted from related prior work [17-18]. Use of a directed CA approach is appropriate when there is scope to extend or complement existing theories around a phenomenon [19], and so suited our explorations of the teams' knowledge sharing behaviors. The directed content analyst approaches the data analysis process using existing theories to identify key concepts and definitions as initial coding categories. In our case, we used theories examining knowledge and behaviors expressed during textual interaction [17-18] to inform our initial categories (Scales 1-8 in Table 3). Should existing theories prove insufficient to capture all relevant insights during preliminary coding, new categories and subcategories should be created [19]. In our context, both authors of this work and two other trained coders first randomly categorized 5% of the communications in a preliminary coding phase. Coders were provided with guidelines for administering, scoring, and interpreting the coding scheme, including examples of messages that were coded under each category. During the pilot coding exercise we found that some aspects of Jazz team members' utterances were not captured by the first version of our protocol (e.g., Instructions and Gratitude) and practitioners in Jazz communicated multiple ideas in their messages. Thus, we segmented the communication using the sentence as the unit of analysis. We extended the protocol by deriving new scales directly from the pilot Jazz data (see scales 9 to 13 in Table 3), after which the first author and the two trained coders recoded all of the messages (1072 messages, noted in Table 1). Duplicate codes were assigned to utterances that demonstrated multiple forms of collaboration, and all coding differences were discussed and resolved by consensus (see Section 4.2 for details). We achieved an 81% inter-rater agreement between the three coders using Holsti's coefficient of reliability measurement (C.R) [20]. This is considered to represent excellent agreement between coders.

**Table 2. Summary Linguistic Measures**

| Linguistic Category | Abbreviation (Abbrev.) | Examples | Reason for Inclusion |
|---|---|---|---|
| Pronouns | I | I, me, mine, my, I'll, I've, myself, I'm | Elevated use of first person plural pronouns (we) is evident during shared situations, whereas, relatively high use of self references (I) has been linked to individualistic attitudes [21]. Use of the second person pronoun (you) may signal the degree to which members rely on (or delegate to) other team members [22]. |
| | we | we, us, our, we've, lets, we'd, we're, we'll | |
| | you | you, your, you'll, you've, y'all, you'd, yours, you're | |
| Cognitive language | insight | think, consider, determined, idea | Software teams were previously found to be most successful when many group members were highly cognitive and were natural solution providers [23]. These traits are also linked to effective task analysis and brainstorming capabilities. |
| | discrep | should, prefer, needed, regardless | |
| | tentat | maybe, perhaps, chance, hopeful | |
| | certain | definitely, always, extremely, certain | |
| Work and Achievement related language | work | feedback, goal, boss, overtime, program, delegate, duty, meeting | Individuals most concerned with task completion and achievement are said to reflect these traits during their communication. Such individuals are most concerned with task success, contributing and initiating ideas and knowledge towards task completion [24]. |
| | achieve | accomplish, attain, closure, resolve, obtain, finalize, fulfill, overcome, solve | |
| Leisure, social and positive language | leisure | club, movie, entertain, gym, party, jog, film | Individuals that are personal and social in nature are said to communicate positive emotion and social words and this trait is said to contribute towards an optimistic group climate [18, 24]. Leisure related language may also be an indicator of a team-friendly atmosphere. |
| | social | give, buddy, love, explain, friend | |
| | posemo | beautiful, relax, perfect, glad, proud | |
| Negative language | negemo | afraid, bitch, hate, suck, dislike, shock, sorry, stupid, terrified | Negative emotion may affect team cohesiveness and group climate. This form of language shows discontent and resentment [25]. |

**Table 3. Coding Categories for Measuring Interaction**

| Scale | Category | Characteristics and Example |
|---|---|---|
| 1 | Type I Question | Ask for information or requesting an answer – "Where should I start looking for the bug?" |
| 2 | Type II Question | Inquire, start a dialogue - "Shall we integrate the new feature into the current iteration, given the approaching deadline?" |
| 3 | Answer | Provide answer for information seeking questions - "The bug was noticed after integrating code change 305, you should start debugging here." |
| 4 | Information sharing | Share information – "Just for your information, we successfully integrated change 305 last evening." |
| 5 | Discussion | Elaborate, exchange, and express ideas or thoughts – "What is most intriguing in re-integrating this feature is how refactoring reveals issues even when no functional changes are made." |
| 6 | Comment | Judgemental – "I disagree that refactoring may be considered the ultimate test of code quality." |
| 7 | Reflection | Evaluation, self-appraisal of experience – "I found solving the problems in change 305 to be exhausting, but I learnt a few techniques that should be useful in the future." |
| 8 | Scaffolding | Provide guidance and suggestions to others – "Let's document the procedures that were involved in solving this problem 305, it may be quite useful." |
| 9 | Instruction/Command | Directive – "Solve task 234 in this iteration, there is quite a bit planned for the next." |
| 10 | Gratitude/Praise | Thankful or offering commendation – "Thanks for your suggestions, your advice actually worked." |
| 11 | Off task | Communication not related to solving the task under consideration – "How was your weekend?" |
| 12 | Apology | Expressing regret or remorse – "Sorry for the oversight and the failure this has caused." |
| 13 | Not Coded | Communication that does not fit codes 1 to 12. |

## 4. RESULTS AND ANALYSIS

Table 4 shows that, of the two projects, a higher number of messages were typically exchanged by those working on user experience tasks. However, those working on the project management project tended to address a higher number of tasks.

**Table 4. Project Measures**

| Project | Messages/Task | Tasks/Contributor | Messages/Contributor |
|---|---|---|---|
| UE | 8.5 | 1.6 | 13.9 |
| PM | 2.9 | 2.3 | 6.8 |

### 4.1 Linguistic Analysis – Attitudes

We examined the paired distributions for each of the linguistic dimensions for normality using the Kolmogorov-Smirnov test and found that there was violation of normality for each. We therefore conducted Mann-Whitney U tests to reveal differences between project type pairs for all the linguistic dimensions in Table 2; these results are presented in Table 5. Here it is evident that the two teams studied expressed significantly different attitudes while solving their respective software tasks ($P<0.05$). These differences were pronounced for all the linguistic dimensions considered.

The results show that those working on the project management tasks were more individualistic (used more "I", "me", "my"), but also that these contributors had more extensive collective team processes (use of more "we", "our", "us" words). On the other hand, contributors to the user experience tasks relied on each other more (used higher amounts of "you", "your", "you're"). For the cognitive dimensions (insight, discrep, tentat, certain), Table 5 shows that practitioners working on project management tasks communicated with significantly more of these processes. Those working around project management tasks were also more work- ("feedback", "goal", "delegate") and achievement-focused (they used higher amounts of "accomplish", "attain", "resolve"). They were also typically more social (communicating with more "give", "buddy", "love"), and spent more time engaging about leisure (e.g., "club", "movie", "party"). They also used more negative language (e.g., "afraid", "hate", "dislike"). In contrast, practitioners working on user experience tasks used significantly more positive language (e.g., "beautiful", "relax", "perfect").

## 4.2 Directed Content Analysis – Knowledge

From the 1072 messages (Table 1), 3448 utterances (sentences) were coded across the two projects. Although contributors to the project management project typically communicated less (as seen in Table 4), our directed CA found that these individuals said more in each message (see Table 6). We present the aggregated interaction behaviors that occurred in the two projects in Table 6. Here it is seen that Information sharing, Discussion, Scaffolding and Comments were the most dominant communication behaviors evident in both sets of contributors' discourses. Table 6 shows that while the other dimensions were less evident, Apology type communication was rarely observed, and only a few utterances were not matched to a category (Not Coded).

**Table 5. Results for Linguistics Analysis**

| Linguistic Category | Abbrev. | UE (Mean Rank) | PM (Mean Rank) | Mann-Whitney Test (*p*-value) |
|---|---|---|---|---|
| Pronouns | I | 511.59 | 555.22 | 0.00 |
|  | we | 503.25 | 561.49 | 0.00 |
|  | you | 552.45 | 524.51 | 0.02 |
| Cognitive | insight | 495.56 | 567.27 | 0.00 |
|  | discrep | 502.77 | 561.85 | 0.00 |
|  | tentat | 497.27 | 565.98 | 0.00 |
|  | certain | 502.40 | 562.13 | 0.00 |
| Work and Achievement | work | 484.01 | 575.95 | 0.00 |
|  | achieve | 477.45 | 580.89 | 0.00 |
| Leisure, social and positive | leisure | 492.17 | 569.82 | 0.00 |
|  | social | 494.65 | 567.96 | 0.00 |
|  | posemo | 631.34 | 465.21 | 0.00 |
| Negative | negemo | 513.11 | 554.08 | 0.00 |

We present the interaction categories as percentages of the overall projects' interactions in order to make comparisons across the two projects. In Table 6 we observe that there was a slightly higher proportion of Type I Questions asked in the user experience project (2.92%) as well as a higher proportion of Answers provided (6.52%) by those working in this project, compared to those working on the project management project (1.84% and 5.08%, respectively). In contrast, those working on the project management project shared more Information (49.32% versus 40.43%). Discussion type utterances were also proportionally more common among those working on the project management project (14.10% compared to 11.07%); however, these results are reversed for Comments (8.76% compared to 5.08%). Findings for Reflection and Scaffolding slightly favored those on the project management project (2.89% and 9.94% versus 2.66% and 8.84%, respectively), while Table 6 shows that those working on the user experience project were Instructed nearly three times as often as those on the project management project (7.55% against 2.76%); a similar pattern is also seen for Gratitude (4.38% compared to 1.66%). A Chi-square test confirmed that there were significant differences in the way practitioners interacted and shared knowledge in the two projects ($X^2 = 109.244$, df = 12, P = 0.000).

**Table 6. Results for Aggregated Interaction Behaviors**

| Category | ∑ codes | | % of utterances | |
|---|---|---|---|---|
|  | UE | PM | UE | PM |
| Type I Question | 34 | 42 | 2.92 | 1.84 |
| Type II Question | 57 | 120 | 4.89 | 5.26 |
| Answer | 76 | 116 | 6.52 | 5.08 |
| Information sharing | 471 | 1126 | 40.43 | 49.32 |
| Discussion | 129 | 322 | 11.07 | 14.1 |
| Comment | 102 | 116 | 8.76 | 5.08 |
| Reflection | 31 | 66 | 2.66 | 2.89 |
| Scaffolding | 103 | 227 | 8.84 | 9.94 |
| Instruction/ Command | 88 | 63 | 7.55 | 2.76 |
| Gratitude/ Praise | 51 | 38 | 4.38 | 1.66 |
| Off task | 20 | 31 | 1.72 | 1.36 |
| Apology | 2 | 11 | 0.17 | 0.48 |
| Not Coded | 1 | 5 | 0.09 | 0.22 |
| ∑ | 1165 | 2283 | - | - |

## 4.3 Attitudes and Knowledge Sharing

We conducted Spearman's Rank Order correlation tests to examine the relationship between our linguistic and directed CA results, as guided by our theoretical justifications in Table 2. For example, we checked to see if levels of reliance language evident in the LIWC analysis was correlated with the number of questions found via the directed CA. During our analysis we observed that tasks in the user experience project were solved in four iterations, whereas those in the project management project were solved over 16 iterations. We therefore divided each project into four equal quarters to ascertain if the attitudes practitioners displayed during their projects corresponded with their knowledge-sharing behaviors. We provide a summary of our results in Table 7. Here it is shown that when practitioners used individualistic language (e.g., "I", "me", "my") they did not provide Answers. However, when there was higher use of collective language (e.g., "we", "our", "us"), practitioners were also involved in Scaffolding; this relationship is strong and statistically significant (P<0.05). In Table 7 it is also shown that when there was high reliance language use (e.g., "you", "your", "you're") there were more Questions. Similarly, when there were higher levels of insightful language (e.g., "think", "believe", "consider") practitioners were also involved with more Scaffolding, and the more work-related terms practitioners used (e.g., "feedback", "goal", "delegate"), the more Answers they provided. In contrast, Table 7 shows that when there were few Comments practitioners used more social language (e.g., "give", "buddy", "love"); we note a strong statistically significant negative correlation for this relationship.

**Table 7. Spearman's Rank Order Correlation Results**

| Linguistic Abbrev. | Directed CA Category | Correlation Coefficient ($r_s(8)$) | Sig. (2-tailed) ($p$-value) |
|---|---|---|---|
| I | Answer | -0.262 | 0.531 |
| we | Scaffolding | 0.881 | 0.004 |
| you | Type I Question | 0.571 | 0.139 |
| insight | Scaffolding | 0.667 | 0.071 |
| work | Answer | 0.619 | 0.102 |
| social | Comment | -0.738 | 0.037 |
| negemo | Discussion | 0.143 | 0.736 |

## 5. COMPARISON AND IMPLICATIONS

The findings uncovered through the adoption of deeper analysis techniques in this work show that studying team behaviors using these methods, even at the word usage level (via LIWC), can lead to enhanced understanding of SE teams. We also believe that outcomes of such deeper enquiries would support critical team strategies [12]. In fact, our preliminary findings offer support for those established by early role theorists who asserted that various behaviors and traits are prevalent and necessary in some team environments or contexts, while other settings demand different attitudes and behaviors for teams to succeed [9]. For example, our use of linguistic analysis has provided insights into the differences in attitudes expressed by two Jazz teams. Of the two teams studied, we found individuals working around project management tasks to be more cognitive and work focused (see Table 5). We compared the teams' memberships to see if our findings were possibly mitigated by the role distribution of these teams and noted that the user experience team comprised 22.32% more programmers and 6.67% more team leaders than the project management team, and the project management team comprised 29.19% more practitioners working across multiple roles; all other roles were similarly distributed. From this preliminary observation we can infer that new team members with relatively lower perceptive abilities may face challenges settling into the project management team. Perhaps the differences noted may be linked to the distribution of roles in these teams, and particularly the dissimilarity for those occupying multiple roles (such individuals are often perceived as highly skilled). These results may be readily validated by studying a larger sample of Jazz projects, and then still further by examining similarly categorized projects in other closed-source case organizations (e.g., Microsoft).

Our directed CA results support more concrete deductions and triangulate previous findings [19]. For instance, as with the linguistic findings for cognitive language and work focus, our directed CA findings uncovered that practitioners working on project management tasks shared higher amounts of information, ideas and suggestions, while those working on the user experience tasks were instructed much more (see Table 6). These findings endorse our proposition of the likely need for more perceptive team members should they be involved in a project management or similar team. Additionally, in terms of project governance, a slightly more critical observation from the directed CA outcomes relates to the much higher level of instructions conveyed to those working on user experience tasks. The need to be instructed may have a negative effect on team leaders' availability, particularly if working in a distributed development context.

Furthermore, when the linguistic and directed CA results were correlated, we uncovered that when there were higher collective processes, practitioners provided more suggestions (Scaffolding), and when there were elevated levels of judgmental attitudes (Comments) practitioners were less social (see Table 7). These intricate details about Jazz teams' dynamics could only be revealed through the adoption of more contextual analysis techniques [3]. From a project governance perspective, these preliminary results suggest that a strategy (and tool mechanisms) that promotes higher team and social focus may be beneficial for task performance – as stressed by agile proponents. A further stage of this study, were it to be feasible, would be to conduct interviews with a sample of these practitioners to supplement our understandings into the specific task qualities (and possible role arrangement) that drive these teams' attitudes and behaviors. We may also enhance these preliminary findings by examining our results against practitioners' actual involvement in software tasks.

Of final note is the balance provided by the approaches used in this work. Here we overcome the limitations of purely quantitative approaches that ignore the complexities of human psychology, and the time-intensive and potentially invasive nature of field work required in full case studies. We encourage those exploring SE team issues to triangulate their frequency-based approaches with contextual analysis techniques.

## 6. LIMITATIONS

Although our primary intent in this study was to assess and compare the outcomes of deeper contextual analysis techniques and to identify opportunities for triangulation, our results may also inform software project governance. Accordingly, we consider possible threats to the validity of our findings:

*Measurement Validity*: The LIWC language constructs used to measure team attitudes have been used previously to study this subject, and were assessed for validity and reliability [15]. However, the adequacy of these constructs in the specific context of software development warrants further investigation. Additionally, our contextual directed CA involving interpretation of textual data is subjective, and so questions naturally arise regarding the validity and reliability of the study outcomes. We employed multiple techniques to deal with these issues. First, our protocol was adapted from those previously employed and tested in the study of interaction and knowledge sharing [17-18], and so there is a strong theoretical basis for its use. Second, we piloted the protocol and extended our instrument by deriving additional codes directly from the Jazz data, and we tested this extended protocol for accuracy, precision and objectivity, receiving an inter-rater measure indicative of excellent agreement [19].

*Internal and External Validity*: The tasks, history logs and messages from the two projects (out of 94) may not represent all the teams' processes in the repository. Work processes and work culture at IBM are also specific to that organization and may not be representative of organization dynamics elsewhere.

## 7. ACKNOWLEDGMENTS

We thank IBM for granting us access to the Jazz repository. Thanks also to the coders for their help. S. Licorish is supported by an AUT VC Doctoral Scholarship Award.